\documentclass[prc,preprint]{revtex4}
\usepackage{graphicx,latexsym}
\usepackage{dcolumn}

\begin{document}

\title{A maximum likelihood method to correct for missed levels based on the $\Delta_3(L)$
statistic.}
\author{Declan Mulhall}
 \affiliation{Department of Physics/Engineering,
 University of Scranton, Scranton, Pennsylvania 18510-4642, USA.}
 \email{mulhalld2@scranton.edu}
\date{\today}
\begin{abstract}
The $\Delta_3(L)$ statistic of Random Matrix Theory is defined as the  average of a set of random numbers $\{ \delta \}$, derived from a spectrum. The distribution $p(\delta)$ of these random numbers is used as the basis of a maximum likelihood method to gauge the fraction $x$ of levels missed in an experimental spectrum. The method is tested on an ensemble of depleted spectra from the gaussian orthogonal ensemble (GOE)
, and accurately returned the correct fraction of missed levels. Neutron resonance data and acoustic spectra of an aluminum block were analyzed. All results were compared with an analysis based on an established  expression for $\Delta_3(L)$ for a depleted GOE spectrum. The effects of intruder levels is examined, and seen to be very similar to that of missed levels. Shell model spectra were seen to give the same $p(\delta)$ as the GOE.
\end{abstract}


\maketitle

\section{\label{sec:intro}Introduction}
Neutron resonance data provide us with a high resolution picture of
the eigenvalues of the nuclear hamiltonian at high excitation
energies.  This  was the birthplace and testing ground for Random
Matrix Theory (RMT) as a model for quantum chaos. For a brief
history of RMT see \cite{guhr}, and for a review of RMT and nuclear
structure, see \cite{weid09}. The correspondence between the
fluctuation properties of nuclear spectra and those of the Gaussian
Orthogonal Ensemble has been verified many times in neutron
resonances \cite{liou72, liou75, jain,frankle94} and proton
resonances \cite{watson81}. Furthermore, shell model calculations
exhibit many of the fluctuation properties of the GOE
\cite{big,horoiran}. For an account of tests of RMT in nuclear
physics see \cite{mitch01}.

The question of the completeness of an experimental spectrum is
important. One needs a gauge of the fraction,  $x$, of the levels
missed in a given experimental spectrum. RMT has already been used
to this end. The fraction of levels not observed due to the finite
resolution and sensitivity of the detectors will change the
distribution of widths from the Porter Thomas distribution which
follows from RMT \cite{brody}. The nearest neighbor distribution
(nnd) is another commonly used statistic. The nnd for a pure
spectrum follows the Wigner distribution,
\begin{equation}
    P(s)=\frac{\pi}{2}se^{-\pi s^2/4},
\end{equation}
where $s=S/D$, $S$ being the spacing between adjacent levels, and $D$ is the average spacing.  The nnd of a spectrum incomplete by a fraction $x$ is given by
\begin{equation}
P(s)=\sum_{k=0}^{\infty} (1-x) x^k P(k;s);\label{eq:pofsx}
\end{equation}
where $P(k;s)$ is the $k^{th}$ nearest neighbor spacing, $E_{k+i}-E_i$. This was first introduced as an ansatz in \cite{watson81}, and rederived in  \cite{agv} and   \cite{bohigas2004}.  Eq.~\ref{eq:pofsx} was used by Agvaanluvsan et al  as the basis for a maximum likelihood method (MLM) to determine $x$ for incomplete spectra \cite{agv}. The $\Delta_3(L)$ statistic (also called the spectral rigidity) introduced by Dyson \cite{dyson} is a commonly used statistic. It is defined as a spectral average:
\begin{eqnarray}
\nonumber \Delta_{3}(L) &=&\left \langle {\rm min}_{A,B}\; \frac{1}{L}\;\int^{E_i+L}_{E_i}dE'\,[\;{\mathcal N}(E')-AE'-B]^{2}\; \;\right\rangle \\
&=&\langle \delta^i_3(L) \rangle,\label{eq:d3}
\end{eqnarray}
where ${\mathcal N}(E)$ is the cumulative level number, the number
of levels with energy $\leq E$ ( its slope is  the level density
$\rho (E)$ ). $A$ and $B$ are chosen to minimize $\delta_3^i(L)$.
They are recalculated for each $i$.  A series of evenly spaced
levels would make ${\mathcal N}(E)$ a regular staircase, and
$\Delta_3(L)=\frac{1}{12}$. At the other extreme, a classically
regular system will lead to a quantum mechanical spectrum with no
level repulsion, the fluctuations will be far greater, and
$\Delta_3(L)=\frac{L}{15}$. The angle brackets mean the average is
to be taken over all positions $E_i$ of the window of length $L$.

An analysis of neutron resonance data using  the $\Delta_3(L)$
statistic and the MLM  of Agvaanluvsan et al was performed in
\cite{mulhall07} and \cite{mulhall09} with  consistent results. When
both methods were tested on ensembles of depleted GOE spectra, the
mean values of $x$ were correct, but the uncertainties were large,
for realistic spectrum sizes.

In this paper we present a new method to test experimental data for missed levels. It is a MLM based on the definition of the $\Delta_3(L)$ statistic. Instead of concentrating on the spectral average of the random numbers $\delta^i_3(L)$ in Eq.~\ref{eq:d3}, we consider instead their distribution. If there are $D$ levels in the spectrum, then, allowing for setting the zero of the energy scale at the lowest level, there are $D-L-1$ values of $\delta_3^i(L)$ for each value of $L$. This amounts to a large sample size of random numbers. In this paper we use numerical simulation to get an expression for the distribution of $\delta_3^i(L)$ for depleted spectra with the fraction missed, $x$, as a parameter, and base a MLM on this distribution. The method will return a most likely value of $x$ for each $L$.

In \cite{bohigas2004}  Bohigas and Pato gave an  expression is given for $\Delta_3(L)$ statistic for incomplete spectra. The fraction of missed levels $x$ is both a scaling factor and a weighting factor and $\Delta_3(L,x)$ is the sum of the GOE and poissonian result:
\begin{equation}\label{eq:bohigas}
    \Delta_3(L,x)=x^2\frac{L/x}{15}+(1-x)^2\Delta_3^{\textrm{GOE}}(L/(1-x)).
\end{equation}
The $\Delta_3(L)$ statistic of an experimental spectrum can be compared with this expression and the best $x$ found. We will see however, that the uncertainty in $x$ is large for this method.

In the next section we describe the process of making, unfolding and depleting GOE spectra, and the calculation of $\delta_3^i(L)$. In Sec. \ref{sec:pd3} we discuss the cumulative distribution function, ${\mathcal N}(\delta)$,  of $\delta_3^i(L)$ for spectra with depletion $x$. A three parameter fit was sufficient for each $x$. The parameters were found for $0.00 \leq x \leq 0.30$ in steps of 0.01. We then fit the parameters as functions of $x$. Now we have ${\mathcal N}(\delta)$ with $x$ as a \textit{continuous parameter}. This is the basis for our MLM. In Sec. \ref{sec:mlm}  the MLM is developed and tested on ensembles of depleted GOE spectra. In Sec. \ref{sec:bohig} we tested Eq.~\ref{eq:bohigas} and used it to return $x$ for depleted GOE spectra, and compared these results with the MLM. In Sec. \ref{sec:data} the method is applied to neutron resonance data and ultrasonic spectra measured from an aluminum block. The results are compared with previous investigations. The effect of intruder levels on $\Delta_3(L)$  and $p(\delta)$ is very similar to missed levels. This issue is addressed in Sec. \ref{sec:intruder}. Recent developments \cite{Koehler2010} questioned the validity of using RMT to model fluctuations of complex spectra. To see if ${\mathcal N}(\delta)$ could discriminate between the GOE and more  physical models  we looked, in Sec.\ref{sec:shell}, at shell model spectra. In Sec. \ref{sec:conc} we make some concluding remarks.

\section{\label{sec:rmt} RMT calculations}
To do our RMT analysis we need to generate an ensemble of random matrices, diagonalize them, unfold them, deplete them, and calculate $\delta_3^i(L)$ for each of them. We will give a brief description of this process here, it is described in morbid detail in \cite{mulhall07}.

The appropriate ensemble for this analysis is the Gaussian Orthogonal Ensemble (GOE) as it describes real, time-reversal-invariant systems. This ensemble is the set of random matrices $H$ whose elements are normally distributed matrix elements, $H_{ij}$, having
\begin{eqnarray}
\nonumber P(H_{i\neq j})=\frac{1}{\sqrt{2 \pi
\sigma^2}}\,e^{-\frac{H_{ij}^2}{2 \sigma^2}}, \quad
P(H_{ii})=\frac{1}{\sqrt{4 \pi
\sigma^2}}\,e^{-\frac{H_{ii}^2}{4 \sigma^2}}
\end{eqnarray}
for the off-diagonal and diagonal elements respectively. The width of the distribution is arbitrary, and we choose $\sigma=1$.  Each of these matrices is diagonalized. We made an ensemble of 3000 matrices, each of dimension $D=3000$. Each matrix has an approximately semicircular level density, with $\rho(E)=\sqrt{4N-E^2},$ for $|E|\leq 2\sqrt{N}$ and $\rho(E)=0$ otherwise. Interesting as this may be, it is not germane to our analysis. The currency of RMT is fluctuations, and in order to compare GOE results with experimental data we must remove this long range (secular) structure from all spectra, including the experimental data, with a process called unfolding. The basic idea is to rescale the energy axis  to give a uniform level density of one level per unit energy, on average. To work with smaller spectra we just take a section from the large ($D=3000$) spectra of whatever size we want.

In a spectrum with picket fence of levels, spaced 1 unit apart, like the harmonic oscillator spectrum, ${\mathcal N}(E)$ is a  staircase with steps 1 unit high and 1 unit long, and the spectrum is said to be rigid. An arbitrary spectrum has ${\mathcal N}(E)=i, E_i \leq E < E_{i+1}$. The $\Delta_3(L)$ statistic is a measure of  fluctuations of ${\mathcal N}(E)$ from a regular staircase, and its definition is the square of the difference between this stairs and a straight line. In the harmonic oscillator case Eq. \ref{eq:d3} can be integrated directly to get $\Delta_3(L) = 1/12$. The situation is messier for an  arbitrary spectrum. Using ${\mathcal N}(E)=i, E_i \leq E < E_{i+1}$, in
Eq. (\ref{eq:d3}), and performing the integral between two
adjacent levels, we come to
\begin{eqnarray}
\nonumber\Delta_3^i(L)=
\frac{1}{L}\;\sum^{i+L-1}_{j=i}\int^{E_j+1}_{E_j}dE'\,(j-AE'-B)^{2}\\
\nonumber= \frac{1}{L}\times(C+VA^2+WA+XAB+YB+ZB^2),
\label{eq:d3iexplicit}
\end{eqnarray}
where $C = \sum^{i+L-1}_{j=i}j^2(E_{j+1}-E_j),\; V =
\frac{1}{3}(E_{i+L}^3-E_i^3),\; W =
\sum^{i+L-1}_{j=i}-j(E_{j+1}^2-E_j^2),\; X =
(E_{i+L}^2-E_i^2),\; Y =
\sum^{i+L-1}_{j=i}-2j(E_{j+1}-E_j),\; Z = (E_{i+L}-E_i)$.
$A$ and $B$ need to minimize $\delta_3^i(L)$, and this leads to the constraints $\partial (\Delta_3^i)/\partial
A = 0$ and $\partial (\Delta_3^i)/\partial B = 0$. These equations readily give $A$ and $B$ that minimize
$\Delta_3^i(L)$ as follows:
\begin{eqnarray}
\nonumber A=\frac{XY-2WZ}{4VZ-X^2}, \quad \nonumber B=
\frac{WX-2VY} {4VZ-X^2}.
\end{eqnarray}

To generate spectra with specific values of $D$ and $x$, we take a section of $\frac{D}{1-x}$ levels from the middle of an unfolded GOE spectra, randomly drop a fraction $x$ of them, and then contract this spectrum by a factor of $1-x$ to restore a level density of 1 level per unit energy. So this gives $D$ levels ``detected" from the original spectrum, and a fraction $x$ ``missed". Note that $x$ is not a continuous parameter. Consider an experimental run of 124 levels. The true spectrum could have 125, 126 \dots or 130 levels,  there being 1, 2, \dots or 6 levels missed, in which case $x$ would have values of 0.8\% , 1.59\% , 2.36\% , 3.13\% , 3.88\%,  or 4.62\%. This should be kept in mind when testing various schemes for estimating $x$.

Now we can make ensembles with a range of $x$ and $D$. We chose to restrict our ensembles to 1500 elements each, and chose   $D=1000$ and $0.00 \leq x \leq 0.30$ in steps of 0.01.

\section{\label{sec:pd3}The distribution of $\delta_3^i(L)$}

Given an unfolded GOE spectrum of size $D$, where a fraction $x$ has been missed, $\Delta_3(L)$ will be the (spectral) average of the set of $D-L-1$ numbers $\delta_3^i(L)$. We would like to get the probability density of these numbers as a function of $x$. In what follows we will write $\delta$ for $\delta_3^i(L)$, dropping all subscripts assuming as fixed value for $L$. So the probability density of $\delta_3^i(L)$ will be written $p(\delta)$, and the cumulative distribution will be written ${\mathcal N}(\delta)$.
\begin{figure}
\includegraphics[width=.4\textheight]{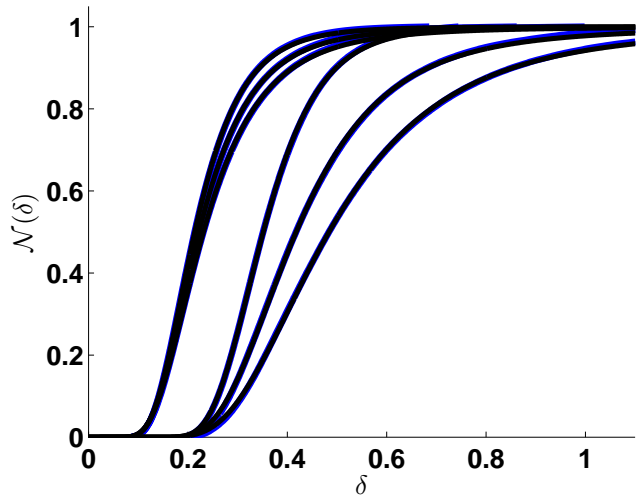}
\caption{\label{fig:cdfd3i}(color online) The cumulative distribution function, ${\mathcal N}(\delta)$. Each set of 3 lines corresponds to, from left to right,  $x=0.00$, 0.04, and 0.08. The leftmost set of 3 lines of lines are for for $L=10$, and the rightmost set have $L=40$. The 3-parameter fit, Eq.\ref{eq:cdf} is plotted in black for each case, and is indistinguishable from the ensemble data (blue).}
\end{figure}

We should explicitly state that this analysis assumes that $p(\delta)$ is ergodic, in that the distribution of $\delta$ is the same for one huge matrix as it is from the superposition of many small ones. It is true that the level spacing distribution $p(s)$  is ergodic. A histogram of the 3000 spacings from a single $D \geq 3001$ spectrum is the same as a histogram of the $s=E_{16}-E_{15}$ spacing of 3000 spectra with $D\geq 16$, for example. But $\Delta_3(L)$ is defined as a spectral average, so while it seems obvious that it is ergodic, it seems prudent to state the assumption explicitly, given that that $\Delta_3(L)$ can be very different from one spectrum to the next see Ref.~\cite{mulhall07}.

We proceeded by guessing the functional form of $p(\delta)$, and were  surprised to see that $p(\delta)$ was a simple function of $\log \delta$ for a wide range of $L$ and $x$, and this motivated us to parameterize the cumulative distribution function.  We used following parameterization:
\begin{equation}\label{eq:cdf}
{\mathcal N}(\delta) = \frac{1}{2}(1 - \textrm{Erf}[a + b \log \delta + c
(\log \delta)^2]).
\end{equation}
This yields, on differentiation
\begin{equation}\label{eq:pd3icdf}
p(\delta)=-\frac{1}{\sqrt{\pi }} \exp{[-\big(a+b\,\log \delta+c
\log \delta^2\big) }^2]\,\big(\frac{b}{\delta}+\frac{2\ c \log \delta}{\delta}\big).
\end{equation}

In Fig.~\ref{fig:cdfd3i} we see the ensemble average of ${\mathcal N}(\delta)$ for $L=10$ and 40, with $x=0.00$, 0.04 and 0.08. The $L=40$ case is more sensitive to depletion than the $L=10$ case, one can clearly see that the spread in the $x=0.00$, 0.04 and 0.08 lines is greater. This is reasonable because the bigger the window size, $L$, more likely it is to fall across the site of a missed level. Consider a spectrum with $D$=1000, and $x=0.02$, there are 20 sites for missed levels. With  $L$ = 10, there are 989 positions for the window, and there are at most, 10 positions, $i$, where $\delta_3^i(L)$ will be different from the $x=0.00$ case, compared to 40 for the $L=40$ case, so low $L$ values give a less sensitive distribution. We note that there are just 3 parameters in the fits, shown in blue in the figure, even so their curves lie on the ensemble average values. The spread of the averaged values for each point in the ${\mathcal N}(\delta)$ graphs was of order $10^{-3}$.

The method used to numerically make the ${\mathcal N}(\delta)$ is best illustrated by the following example for $L$ = 20, and $x=0.03$. Take a GOE spectra,  with $D=3000$. Take the middle 100/0.97=1031 levels to avoid end effects. Unfold it. Randomly drop 31 levels. Contract the spectrum by a factor of $(1-x)$, now the level density is 1. Calculate the set of  1000-20-1=979 numbers $\{\delta\}$. Sort them. Do this 1200 times, and get $\{\bar{\delta}\}$ the average of the sorted sets. The standard deviation of the sets was $\sim 10^{-3}$. Now pair $\bar{\delta}_i$, the $\textrm{i}^{th}$ element of $\{\bar{\delta}\}$, with i/979, to get the set $\{\bar{\delta}_i,i/979\}$. A plot of $\{\bar{\delta}_i,i/979\}$ is a graph of ${\mathcal N}(\delta)$.

\begin{figure}
\includegraphics[width=.6\textheight]{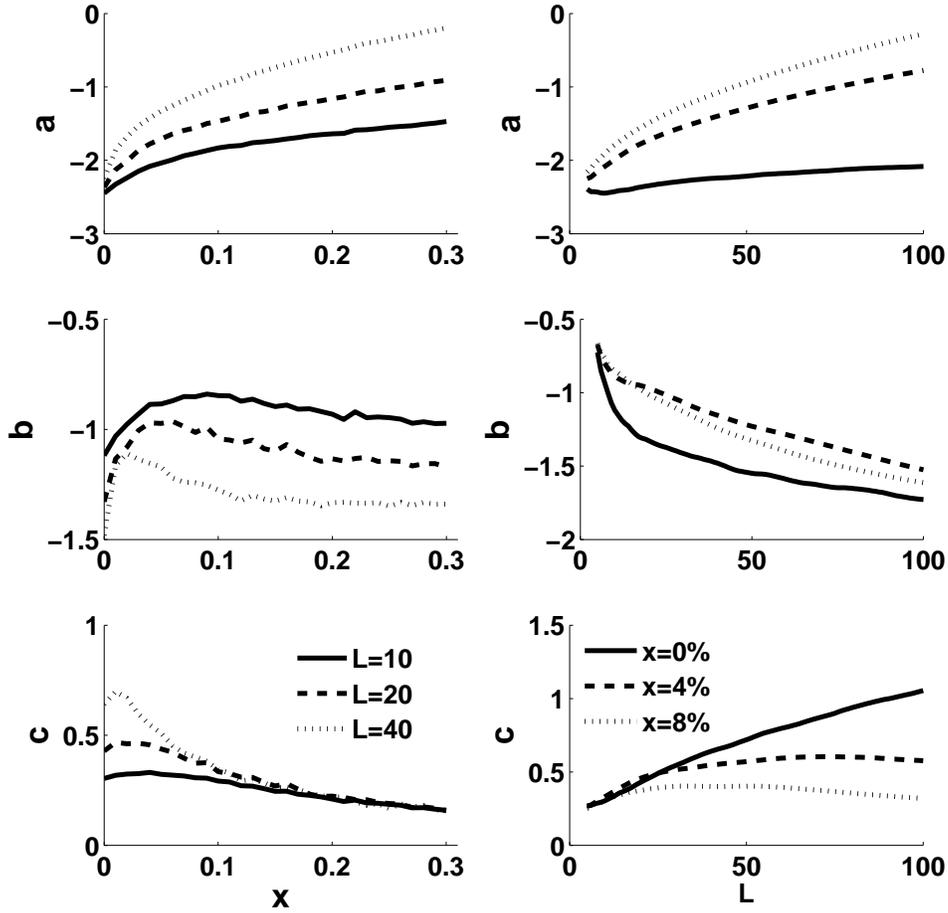}
\caption{\label{fig:abc}The parameters $a$, $b$, and $c$ of  Eq.\ref{eq:cdf}. On the left panels, the parameters are plotted vs $x$ for fixed values of $L$. On the right panels we have plots of the parameters vs $L$ for fixed $x$. From data like this we extracted $a_L(x),\,b_L(x)$ and $c_L(x)$}
\end{figure}

\begin{figure}
\includegraphics[width=.4\textheight]{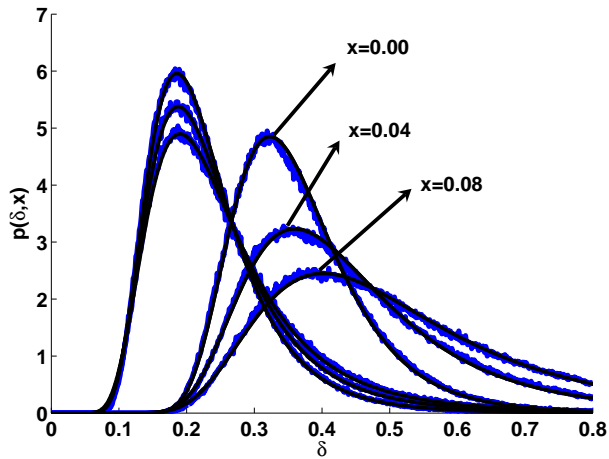}
\caption{\label{fig:pd3i} (color online) The effect of depletion on $p(\delta)$ is shown here, where we have plotted the ensemble average of $p(\delta)$. The set of 3 lines on the left is for $\delta = \delta^i_3(10)$, while  $\delta = \delta^i_3(40)$ is on the right. Each set of 3 lines corresponds to $x=0.00$, 0.04, and 0.08. The parameterized $p(\delta,x)$, Eq.~\ref{eq:plx} is plotted in black , and is barely distinguishable from the ensemble average, in blue. }
\end{figure}

The values of $a$, $b$, and $c$ were obtained  by  fitting Eq.\ref{eq:cdf} to $\{\bar{\delta}_i,i\}$. This was done for $L=5$ to 90 in steps of 5, and for $x$ =0.00 to 0.30 in steps of 0.01. The parameters were smooth values of $x$ for all $L$. See Fig.\ref{fig:abc}. For each value of $L$ the parameters were fit to smooth functions of $x$, $a_L(x)=a_0 + a_{\frac{1}{2}} \sqrt{x}  + a_1 x + a_2 x^2$, with similar expressions for $b_L(x)$ and $c_L(x)$. The values of $a_0 , a_{\frac{1}{2}} , a_1, a_2, \dots, c_1, c_2$ were calculated for all values of $L$ from 5 to 100.  We now have a probability density for $\delta_3^i(L)$ with $x$ as a continuous parameter:
\begin{equation}\label{eq:plx}
p(\delta,x)=-\frac{1}{\sqrt{\pi }} \exp{[-\big(a_L(x)+b_L(x)\,\log \delta+c_L(x)
\log \delta^2\big) }^2]\,\big(\frac{b_L(x)}{\delta}+\frac{2\ c_L(x) \log \delta}{\delta}\big).
\end{equation}
In Fig.\ref{fig:pd3i} we have some examples of $p(\delta,x)$. When the fitted values, $a(x), b(x),$ and $c(x)$ were used in $p(\delta,x)$ the results were indistinguishable from when those values of $a$, $b$, and $c$ that were got from the fitting procedure were used. Again, we anticipate from the graph that higher values of $L$ will be more useful in gauging $x$. In the next section we see how these parameters are used to get $x$ for an unfolded spectrum.

As a test of our machinery, we checked that $p(\delta)$ was independent of $x$ for completely uncorrelated (poissonian) spectra, and it was.

\begin{figure}
\includegraphics[width=.6\textheight]{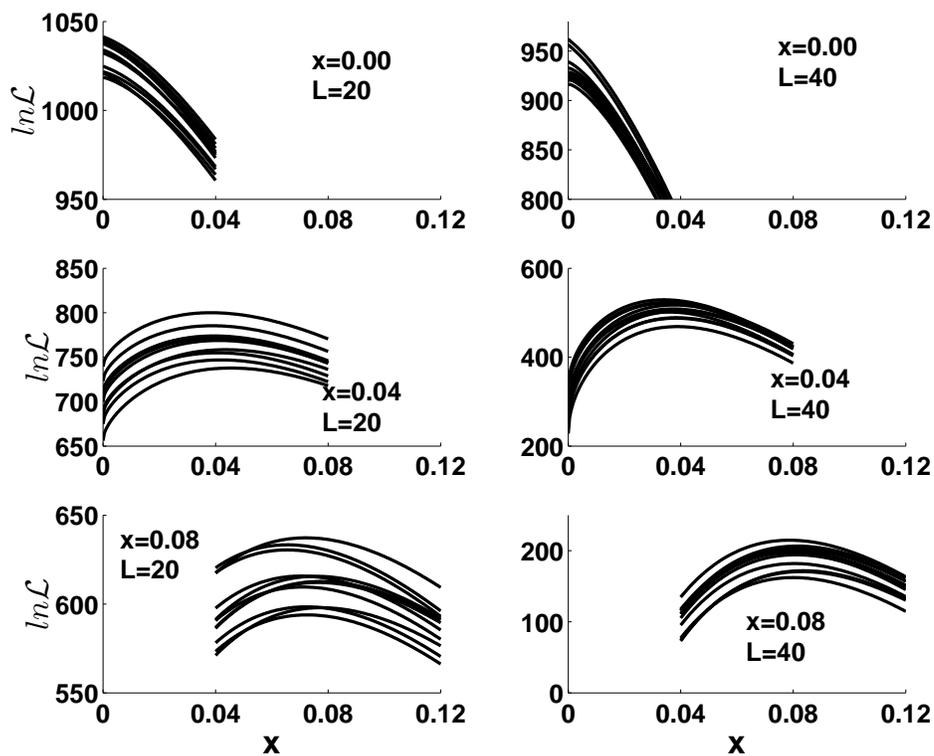}
\caption{\label{fig:mlmL}The results of the MLM calculation for 10 randomly chosen depleted members of the GOE with  $x=0.00$, 0.04, and 0.08. The left hand side has $L=20$, and the right hand side has $L=40$.}
\end{figure}

\begin{figure}
\includegraphics[width=.6\textheight]{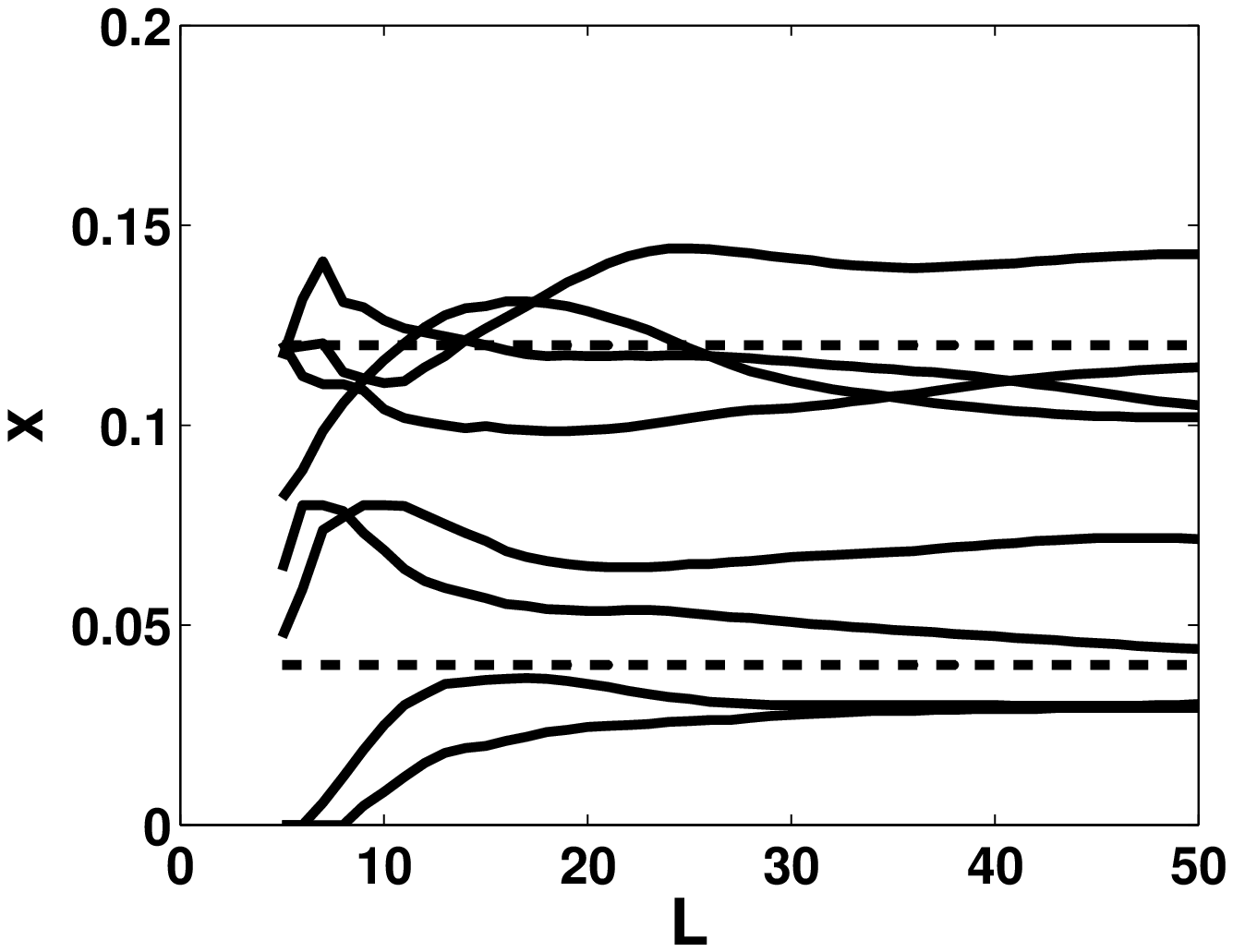}
\caption{\label{fig:mlmx}The results of the MLM calculation for randomly chosen spectra with  $x=0.04$ and 0.12.}
\end{figure}
\section{\label{sec:mlm}The maximum likelihood method. }
\begin{figure}
\includegraphics[width=.6\textheight]{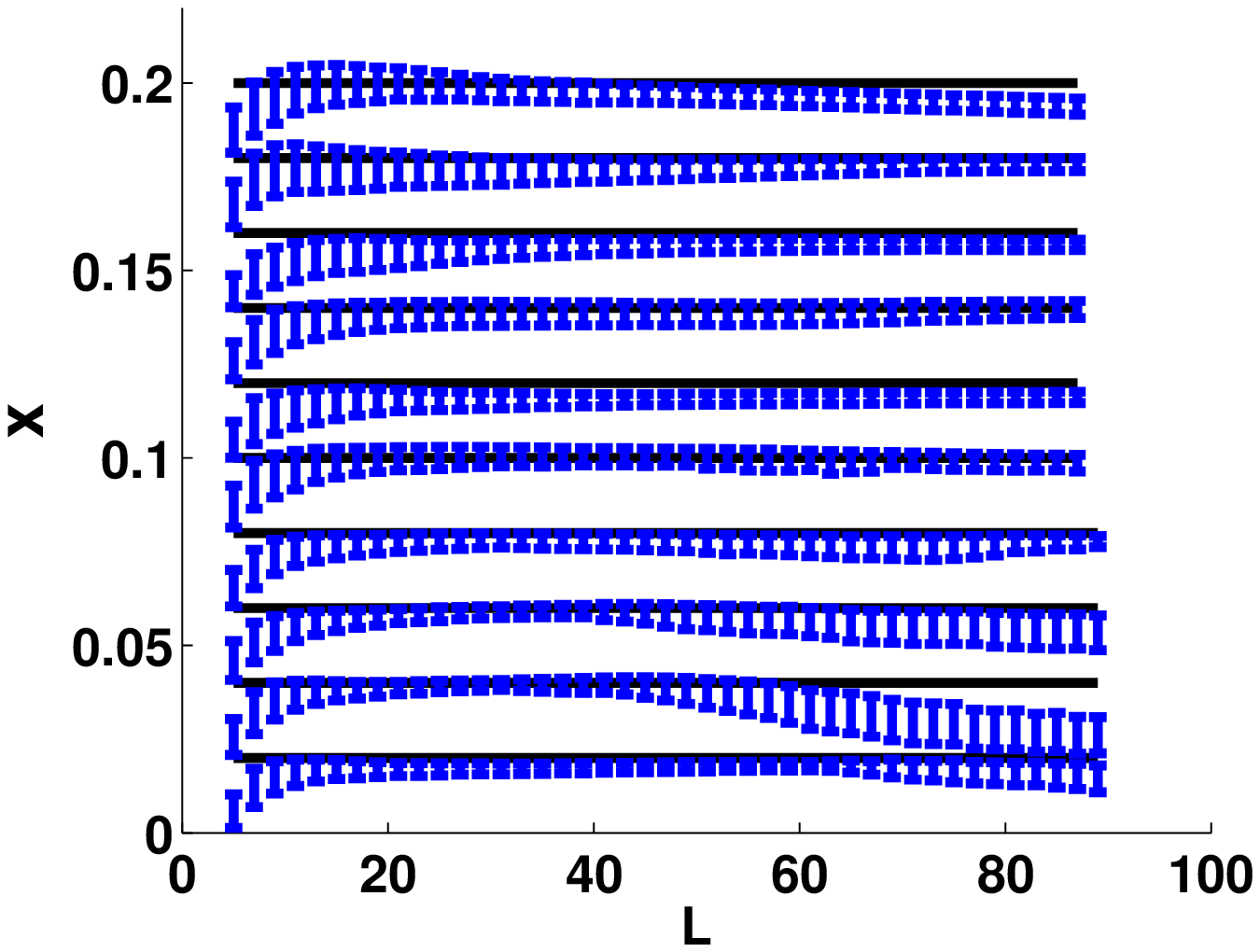}
\caption{\label{fig:mlmAllL}The results of the MLM calculation for $x=0.02$, 0.04, $\dots$  0.18 and 0.20, for all L. $\sigma_x$ was used for errorbars. We used this graph as a guide for using $\sigma_x=0.006$ for $20 \leq L \leq 40$ for all $x$.}
\end{figure}

\begin{figure}
\includegraphics[width=.6\textheight]{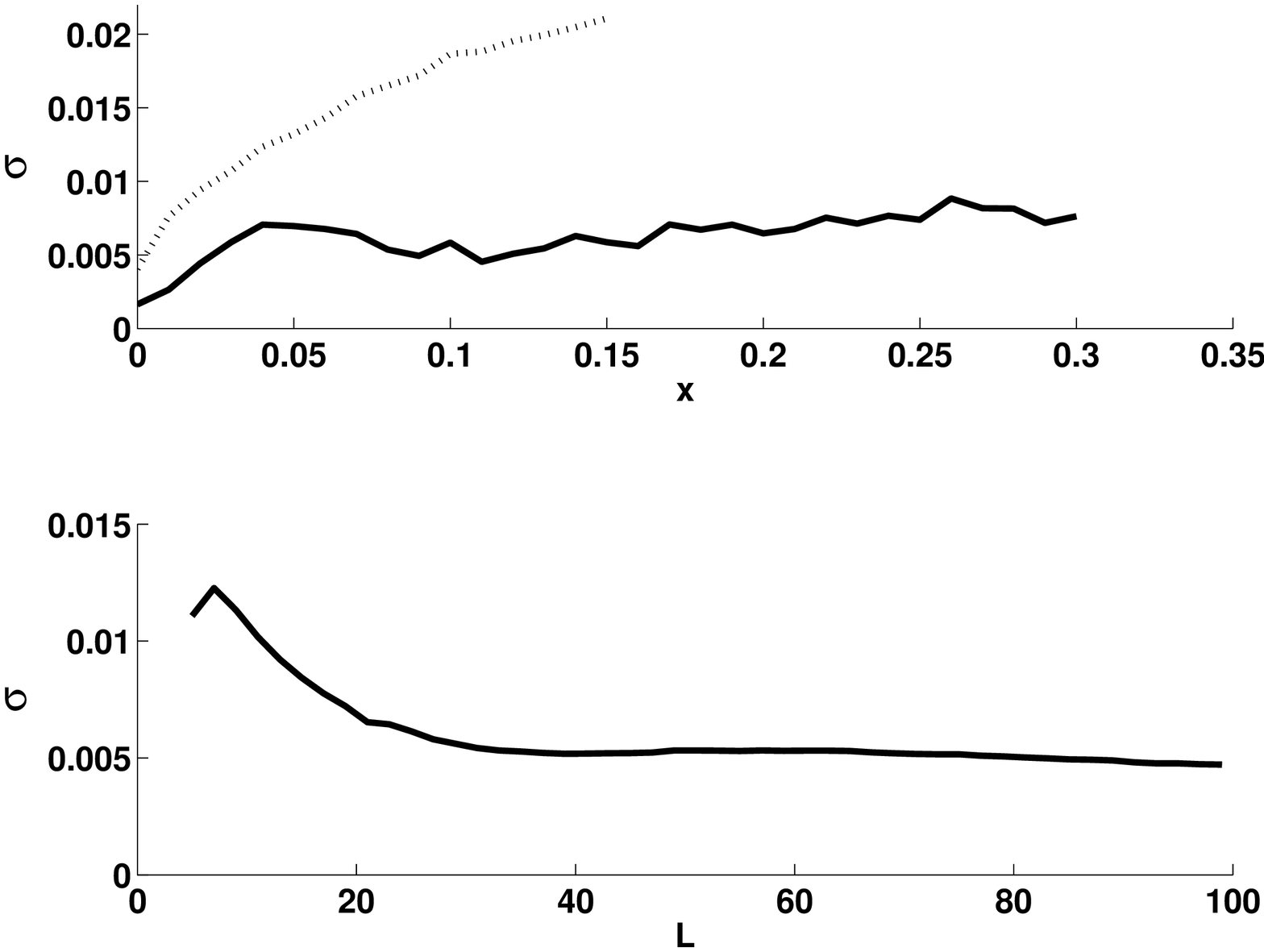} \caption{\label{fig:sigxl} The mean value of $\sigma$ averaged over  $L$, top panel, and $x$, bottom panel. The dashed line in th eiupper panel is the result for the analysis based on Eq.~\ref{eq:bohigas}}
\end{figure}

Now that we have the probability distribution of $\delta_3^i(L)$ parameterized, Eq.~\ref{eq:plx}, we can use it to find the most likely value of $x$: given a set $\{ \delta_3^i(L) \}$ for some $L$, the most likely value of $x$ is the one that maximizes the likelihood ${\mathcal L}=\prod_i p_L(\delta_3^i(L),x)$. In practice we work with $\log {\mathcal L}$. We tested this method on an ensemble of 300 spectra, with $N=1000$ after depletion, with values of $x=0.00, 0.01, 0.02, \dots,0.29, 0.30$. The mean, $\bar{x}$ and standard deviation, $\sigma_x$, of the 300 values of $x$ were returned for $L$. In Fig.~\ref{fig:mlmL} we see some representative  examples of $\log {\mathcal L}$, and in Fig.~\ref{fig:mlmx} we see the results of $\bar{x}$ for 8 spectra, 4 with $x$= 0.04, and 4 with 0.12. Each set of 4 spectra were randomly chosen, but they are representative of the general behavior of $x$ vs $L$. In Fig.~\ref{fig:mlmAllL} we see the ensemble average for $x$=0.04, 0.06, $\dots$ 0.014, 0.016. We used $\sigma_x$ as errorbars in this plot. The figure suggests that the most reliable range of $L$ to use has $20 \leq L \leq 40$, because in this range, $\sigma_x$ settles down to  a smaller value, see Fig.~\ref{fig:sigxl}, and $\bar{x}$ is close to the true value of $x$.

The MLM doesn't have an error bar for the most likely value of $x$ it returns for a specific spectrum. In the analysis of an individual spectrum, one may report a graph of $\log {\mathcal L}$ vs $x$, and state its maximum. If the peak in $\log {\mathcal L}$ is sharp, one would have more confidence in the results. Agvaanluvsan  et al \cite{agv},  used the broadness of the graph of $\log {\mathcal L}$ vs $x$ to give a range for $x$. However, if the spectrum being analyzed is from a known ensemble then $\sigma_x$ as described above would be a reasonable gauge of how close to the true value of $x$ the MLM gets. Looking at $\sigma_x$ averaged over either $x$ or $L$, we are justified in using a value of $\sigma=0.006$ in our analysis, see Fig.~\ref{fig:sigxl}.

Spectrum size is an issue of great practical importance when applying these results. In the case of neutron resonance data, the spectra we analyzed had typically 80 to 90 levels, and we also looked at subsets of them. The acoustic spectra we examined had $\approx 250$ levels. So regardless of the way we choose the error bar, we need to state the $N$ dependance of it. Each spectrum in our test had $N=1000$, and this yielded $N-L-1$ values of $\delta$. We suggest, based on the behavior of $\sigma_x$ that we just calculate $x$ for $10 \leq L \leq 40$, take $\sigma_x \approx 0.006$, and for spectra of size $N$, use $\sigma = 0.006 \, \sqrt{ \frac{1000}{N-L}}$. This is obviously a very rough rule of thumb, don't forget that when $N=80$ the lowest values $x$ can have are 1.25\%, 2.5\%, 3.75\%, and 5\%, corresponding to 1, 2, 3 and 4 levels missed. We will see that the drift in the returned value of $x$ for a given spectrum is often the biggest consideration for extracting $x$.

\section{\label{sec:bohig}The Bohigas expression for $\Delta_3(L,x)$}
\begin{figure}
\includegraphics[width=.6\textheight]{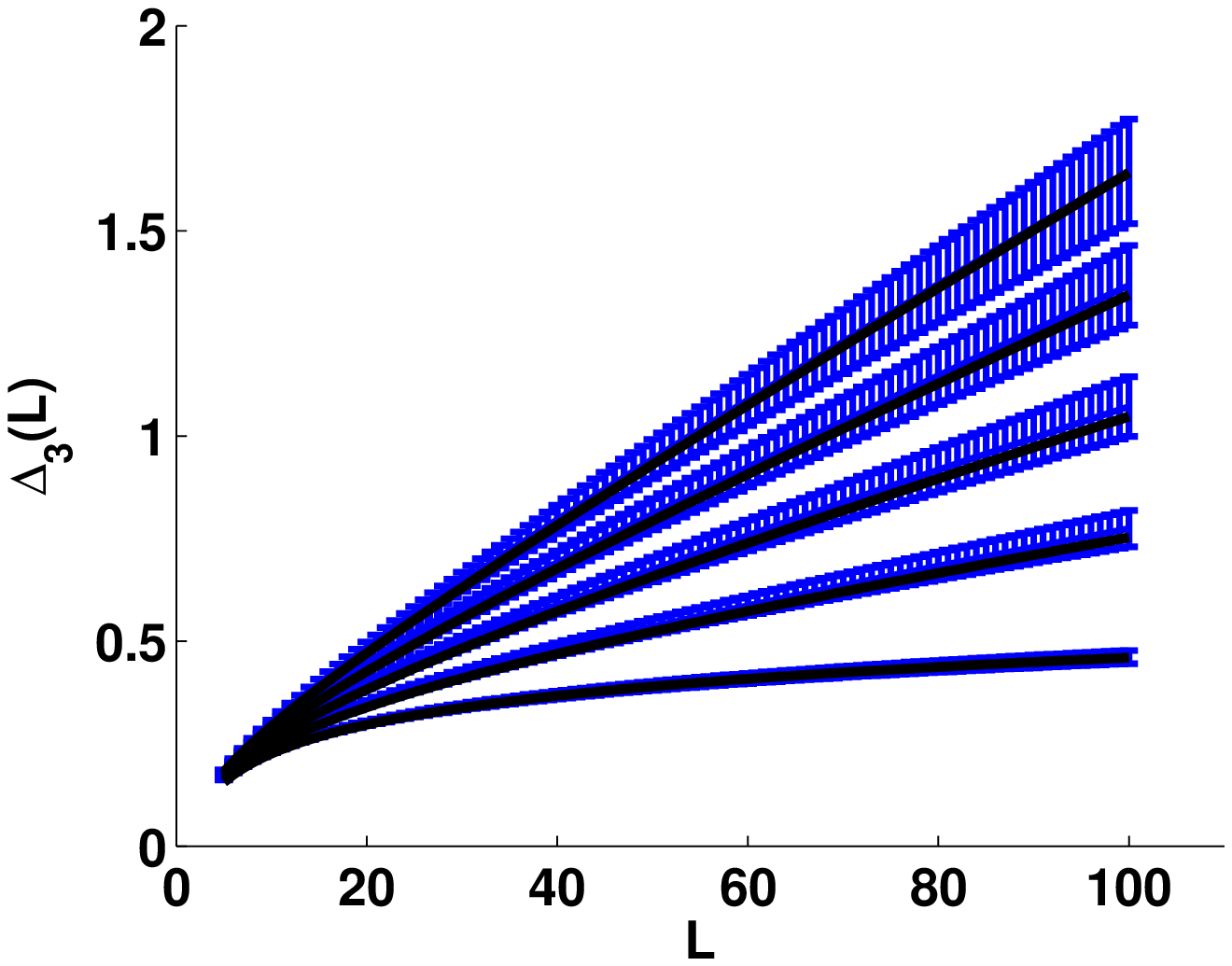}
\caption{\label{fig:bohgoe} A comparison between $\Delta_3(L)$ vs $L$ for the Bohigas expression and the ensemble averages.}
\end{figure}

\begin{figure}
\includegraphics[width=.6\textheight]{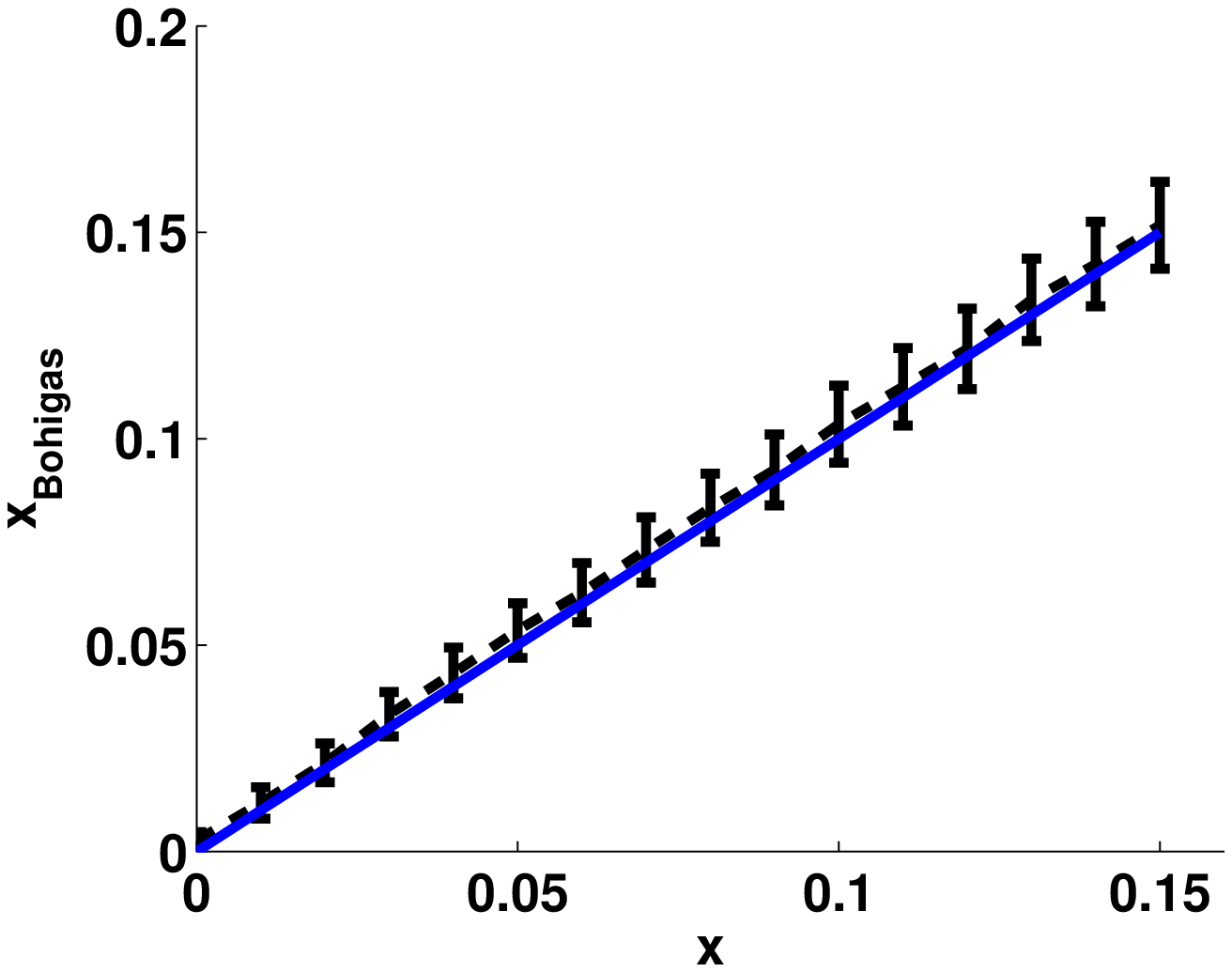}
\caption{\label{fig:bohx} The test results  for using the Bohigas expression to extract $x$.}
\end{figure}

In this section we use Eq.~\ref{eq:bohigas} to extract $x$ from the depleted GOE, and compare the results with that of our MLM. A comparison of $\Delta_3(L)$ for the depleted GOE with Eq.~\ref{eq:bohigas} gives an excellent agreement. In  Fig.~\ref{fig:bohgoe} we see the results for $x=0.00$, 0.05, 0.10, 0.15 and 0.20. The Bohigas result lies very close to the GOE results in blue. To test Eq.~\ref{eq:bohigas} as a tool for gauging  $x$ we get $\Delta_3(L)$  for a depleted spectra, and find the $x$ that minimizes $\sum_L (\Delta_3(L)-\Delta_3(L,x))^2$. Repeating this for 1000 spectra for with  $x$ from 0.00 to 0.15 in steps of 0.01, we got the results are shown in Fig.~\ref{fig:bohx}. The mean value of $x$ was very accurate. The error bars, however, are much bigger than for the MLM. In Fig.~\ref{fig:sigxl} top panel the dashed line is the $\sigma_x$ from this analysis. It is much bigger than $\sigma_x$ for our MLM. In Table~\ref{tab:data} we include a column of results from this method.

\section{\label{sec:data}Application to neutron resonances and ultrasonic spectra.}

In Ref.\cite{mulhall07, mulhall09} the $\Delta_3(L)$ statistic and the MLM of Agvaanuvlusaan et al. was used to to gauge the completeness of neutron resonance data. The results from both methods were consistent with each other. The uncertainties in $x$ for each of these methods was around 0.03. Here we do a new analysis of some of the same datasets with the new mlm, and report the results. A summary of the data sets is in Table \ref{tab:data}. We analyzed neutron resonance data from 7 isotopes in all. The data was taken from the Los Alamos National Laboratory website \footnote{http://t2.lanl.gov/cgi-bin/nuclides/endind.}. Some of the spectra examined did not yield a flat $x$ vs $L$ graph. In these cases the average $x$ is meaningless, and we report that the method is inconclusive. When $x$ vs $L$ is flat, we report the result of the MLM as  $x=\bar{x}\pm \sigma_x$, where the average is taken over the range $20 \leq L \leq 40$, and  $\sigma_x=0.006 \sqrt{1000/D-30}$, where $D$ is the number of levels.

The cumulative level number gives the first indication of the purity of the data. Kinks in ${\mathcal N}(E)$ leading to smaller slopes would suggest a section of data where levels were missing. Sometimes data sets were compiled from different laboratories. The nuclear level density is essentially constant in the range of energies of neutron resonance data, so abrupt drops in the level density suggest experimental issues. Using this as a guideline, some data sets were split into subsets.

\subsection{$^{158}$Gd} In panel a) of Fig.~\ref{fig:mlmisotopes} we see the results of the MLM. There were 93 levels in all, and a drop in $\rho(E)$ at the $60^{\textrm{th}}$ level indicated that the lowest 60 levels were a more pure set than the higher 32 levels. The top dashed line, with $x=0.30$ is for the higher 32 levels. In our MLM, the maximum value we went to was $x=0.30$. A spectrum with $x>0.30$ would ideally return a flat line for $x$ vs $L$ at $x=0.30$, suggesting in this case, that there were originally at least 47 levels in this range. The results for the full set (solid line) are  $\bar{x}=0.121 \pm 0.024$ which translates into there being $93/(1-x)=105 \pm 3$ levels initially, $13 \pm 3$ being missed. This is consistent with the lower 60 levels being pure, and the upper 32 levels being 0.78 of the full spectrum in that range.

\subsection{$^{58}$Ni} Guided by ${\mathcal N}(E)$ we took the  the full set of 63 levels for the $^{58}$Ni data. A plot of $x$ vs $L$ shown in Fig.~\ref{fig:mlmisotopes} panel b), raises serious questions about the reliability of the MLM in this case. The results are  inconclusive.

\subsection{$^{152}$Sm} An analysis of the full set of 91 levels gives an $x=0.081 \pm 0.024$, as seen in shown in Fig.~\ref{fig:mlmisotopes} panel c), solid line. This corresponds to there being $99 \pm 3$ levels in the full spectrum, with $8 \pm 3$ missed. The first 70 levels look pure, so it seems that the missed levels were in the upper range.

\subsection{$^{234}$U} The 118 levels in the full set of resonances yielded higher values of $x$ as $L$ increased, solid line in panel d), Fig.~\ref{fig:mlmisotopes}. As in the case of $^{54}$Fe and $^{58}$Ni, any conclusions are therefore suspect. There was a kink in ${\mathcal N}(E)$ after the $78^{\textrm{th}}$ level. These first 78 levels had a monotonic increasing  $x$ vs $L$ curve, while the top 20 levels had a decreasing curve. Little can be concluded from this.

\subsection{$^{236}$U} There were 81 levels in the $^{236}$U set. The full set had $x=0.124 \pm 0.027$ corresponding to $10 \pm 2$ levels missed. Guided by a kink in ${\mathcal N}(E)$ at level 70, we analyzed the lowest 69  levels and got $x=0.038 \pm 0.030$  corresponding to $3 \pm 2$ levels missed in that range.

\subsection{$^{235}$U} 
\begin{figure}
\includegraphics[width=.6\textheight]{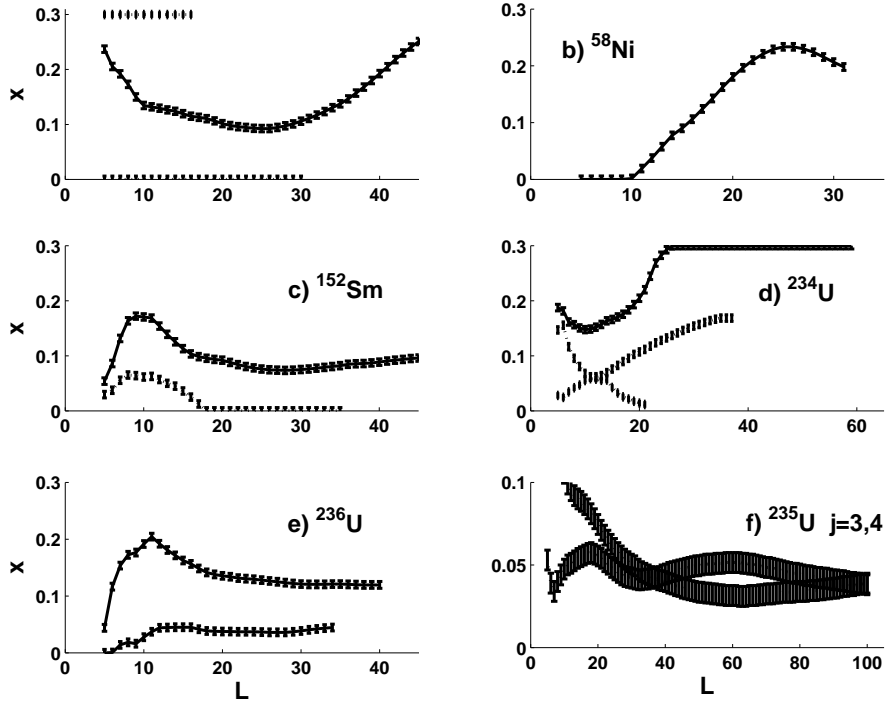} \caption{\label{fig:mlmisotopes}The results of the MLM calculation for neutron resonance data. In panel a) we have $^{158}\textrm{Gd}$. The results for the full set (solid line) lies between those for the subset containing just the lowest 60 levels, (lower dashed line) and the subset containing the top 32 levels (upper line). In panel b) we have  $^{58}\textrm{Ni}$, nothing can be concluded from this about $x$.  In c) we see the full spectrum of $^{152}\textrm{Sm}$ (solid line) above the lowest 70 levels (lower dashed line. In d) we have the full $^{234}\textrm{U}$ data set (solid line) looking more incomplete that the first 75 levels alone, (middle line dashed), while the levels 76 to 118 (lower dashed line) look like a complete subset. In e) we have the full $^{236}\textrm{U}$ set of 81 levels looking more incomplete than the lowest 69 levels. Finally in panel f) we see the $^{235}\textrm{U}$ data. The $j=3$ subset is the solid line, and the dashed line is the $j=4$ subset. Note the range on the vertical axis.}\end{figure}

This is a spectacular data set, with over 3100 levels. Guided by a level density plot, we analyzed the lowest 950 levels. The target is odd, with $j=\frac{7}{2}$, so the neutron resonances were compound states of the $^{236}$U nucleus with $j=3$ and 4. These resonances were labeled with angular momentum, and we separated the 2 sequences of levels and analyzed them separately. The result is in panel f), Fig.~\ref{fig:mlmisotopes}, where the $j=3$ subset is the solid line, with $x=0.053 \pm 0.010$, corresponding to $20 \pm 4$ levels missed. The dashed line is the $j=4$ subset. The mean value of $x$ is 0.029 for the $j=3$ set, and 0.031 for the $j=4$ set, with $x=0.045 \pm 0.008$, corresponding to $26 \pm 5$ levels missed. Note the range of $L$, and how flat the lines are. This result is consistent with the other estimates of $x$ in \cite{mulhall07}.

\subsection{Acoustic data}
\begin{figure}
\includegraphics[width=.6\textheight]{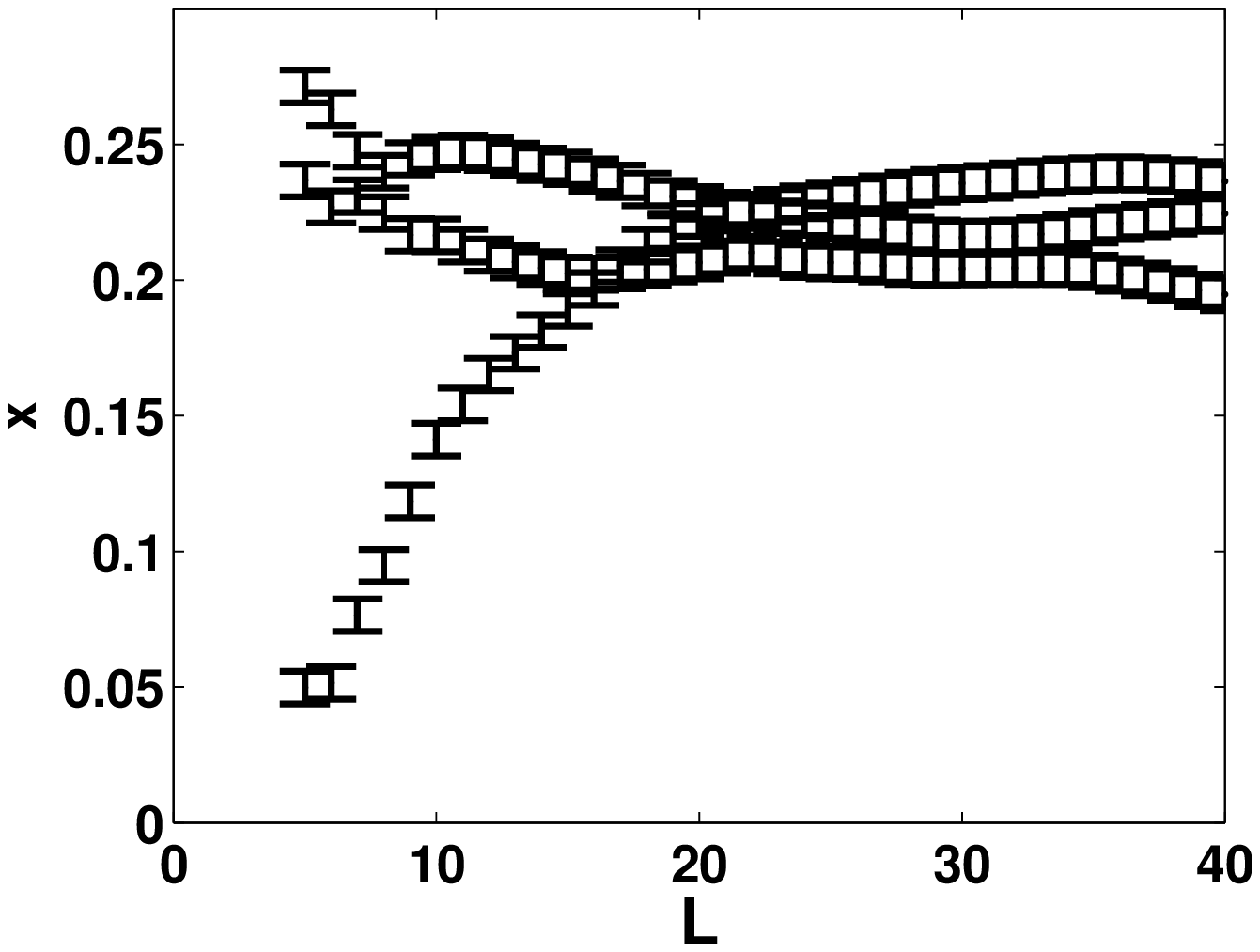}
\caption{\label{fig:oleg}The results of the MLM calculation for acoustic data for sets 1,2, and 4. The other sets had $x> 0.3$.  The mean values for the interval $20 \leq L \leq 40$ are  0.20, 0.22 and 0.23 respectively.}
\end{figure}

Here the spectra were resonant frequencies of aluminum cavities. The full experiment is described in \cite{oleg}. The cavities used in the experiment are made out of aluminum cubes with a cube side size of d=20mm.   The symmetry of the cube is broken by additional features such as an asymmetrically placed cylindrical well and a  removed side corner.  The radius of the well is 5mm and its depth is 18mm. Different configurations of the transducers on the blocks gave different data sets. The experimentalists used a comparison of $\Delta_3(L)$  with depleted GOE results, and conclude that there were 25\% of resonances missed. Only data sets 1,2 and 4 of the 6 data sets analyzed gave $x$ vs $L$ curves less than 0.30. These are plotted in  Fig.~\ref{fig:oleg}. The MLM result for all the data sets from 1 to 6 respectively are  $x=0.204 \pm 0.012$, $x=0.221 \pm 0.014$, $x>0.30$, $x=0.234 \pm 0.014$, $x>0.30$, and $x>0.30$. The corresponding results for the Bohigas method are 0.28, 0.20, 0.4, 0.23, 0.29  and 0.4

\section{\label{sec:intruder}Intruder levels}
\begin{figure}
\includegraphics[width=.6\textheight]{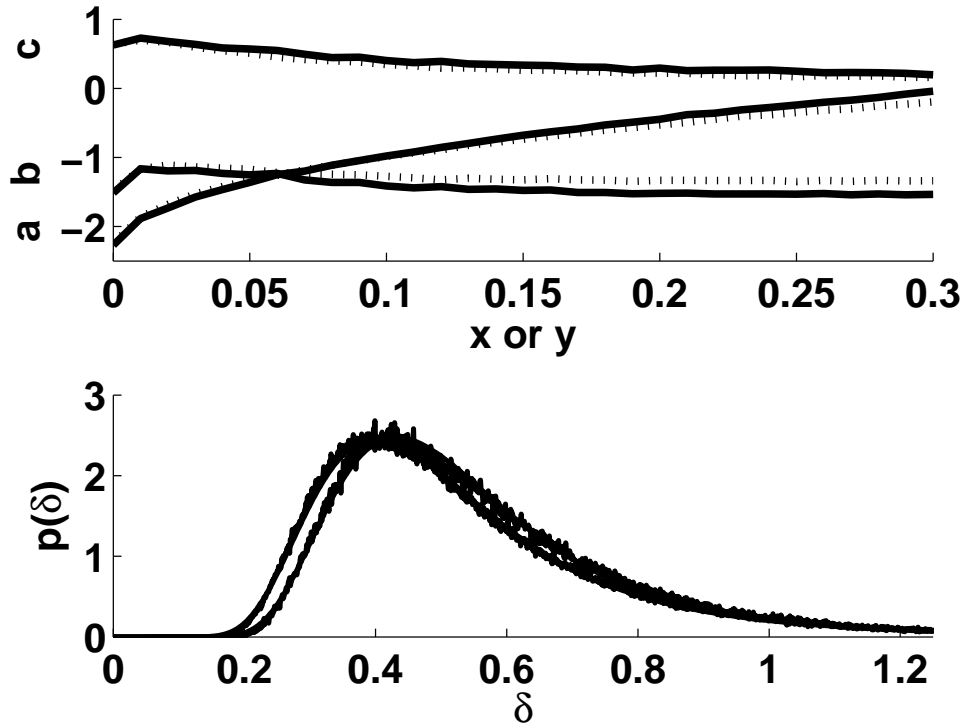}
\caption{\label{fig:d3xy}Lower panel $p(\delta)$ for spectra with x=0.08 depletion (left peak) and y=0.08 intruders. They are quite close together and it is hard to distinguish the effect of missed levels from intruders.  The ensemble average data and the fits are superimposed, and indistinguishable. The upper panel shows the parameters $a,b,$ and $c$ for in the fit Eq.~\ref{eq:plx}. The solid lines are for intruders and the dashed lines are for missed levels.}
\end{figure}

In spectroscopy it is quite possible to falsely label a background noise peak as a level, or to have a level with a different angular momentum to appear. It is important in RMT to know how many sequences of levels are present, (a sequence is a set of energies with the same quantum numbers). If  an intruder from a different sequence is present, it will not be repelled by the other levels, and cause differences in the fluctuation properties of the spectra. In neutron resonance data, $s$-wave neutrons on an even-$A$ target give a set of spin-$\frac{1}{2}$ levels, one could have a spin-$\frac{3}{2}$ level in their midst from a $p$-wave neutron. An examination of ${\mathcal N}(\delta)$ was performed for spectra with a fraction $y$ of intruder levels. An ensemble of 1200 spectra of size $(1-y)D$ were prepared, with D=1000, and were then polluted by adding $y D$ ``levels", which were just random numbers with a uniform distribution over the range of the spectrum. The calculational details were much the same as for depletion. The results were surprisingly similar to those for depletion. In  Fig.~\ref{fig:d3xy}, lower panel, we see $p(\delta)$ for the cases of $x=0.08$ depletion, and $y=0.08$ intruders. In both cases $L=40$. The upper panel shows the parameters in the fit to Eq.~\ref{eq:cdf} for both cases.

\begin{table}
\caption{\label{tab:data}The results for $x$, the percent of
missing levels in the data. }
\begin{ruledtabular}
\begin{tabular}{lllllll}
Isotope & NND & $\Delta_3(L)$(Bohigas) & $p(\delta)$ & N (\# levels) & subset \\
\hline
$^{58}\text{N}$ & 0\% & 18\% & Inconclusive & 63 & All \\
$^{152}\text{Sm}$ & 3\% & 0\% & 0\% & 91 &$1 \rightarrow 70$ \\
$^{152}\text{Sm}$ & 3\% & 10\% & 8\% $\pm$ 2\% & 91 &All \\
$^{158}\text{Gd}$ & 11\% & 13\% & 12\% $\pm$ 2\% & 93 &All \\
$^{158}\text{Gd}$ & 0\% & 0\% & 0\% & 93 &$1 \rightarrow 60$ \\
$^{158}\text{Gd}$ & 12\% & 42\% &$>$30\% & 93 &$61 \rightarrow 93$ \\
$^{234}\text{U}$ & 9\% & 40\% & Inconclusive & 118 & All\\
$^{234}\text{U}$ & 6\% & 13\% & Inconclusive & 118 &$1 \rightarrow 75$ \\
$^{234}\text{U}$ & 7\% & 4\% & Inconclusive & 118 &$76 \rightarrow 118$ \\
$^{236}\text{U}$ & 5\% & 20\% &  12\% $\pm$ 3\%  &81 & All\\
$^{236}\text{U}$ & 0\% & 5\% & 4\% $\pm$ 3\% & 81 &$1 \rightarrow 69$ \\
$^{235}\textmd{U}\quad j=3$ & 3\% & 9\% & 5\% $\pm$ 1\%  & 1436 & $1 \rightarrow 381$ \\
$^{235}\textmd{U}\quad j=4$ & 2\% & 4\% & 5\% $\pm$ 1\%  & 1732 & $1 \rightarrow 569$
\end{tabular}
\end{ruledtabular}
\end{table}

\begin{figure}
\includegraphics[width=.6\textheight]{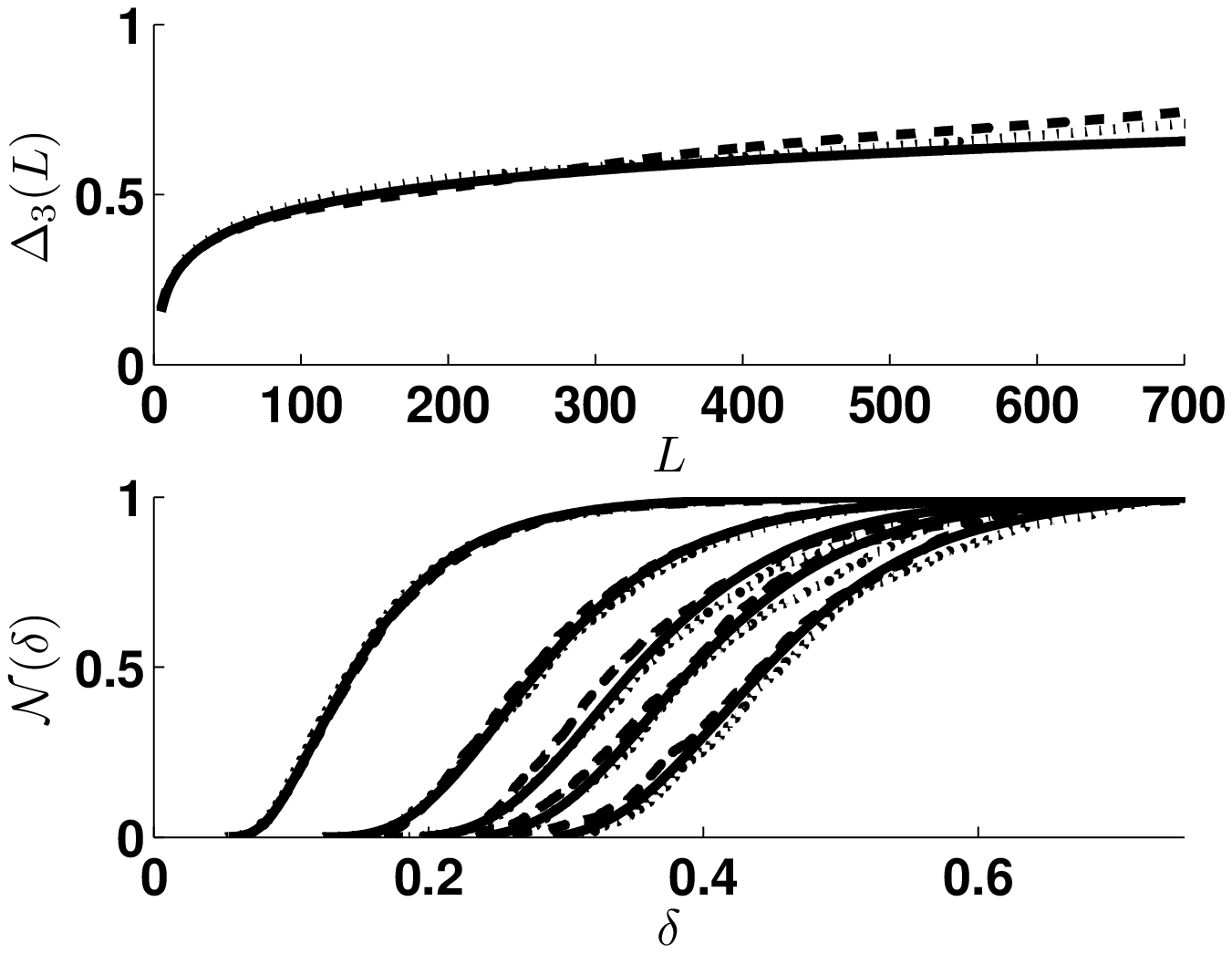}
\caption{\label{fig:shell}The $\Delta_3(L)$ and ${\mathcal N}(\delta)$ for shell model spectra with  $J^+T=2^+1$ (dashed line), and $J^+T=2^+0$ (dotted line), compared to the GOE result (solid line) for pure GOE spectra. }
\end{figure}

\section{\label{sec:shell}Shell model spectra}
Recent developments have called into question the validity of
applying RMT to describe the fluctuation properties of complex
spectra. In \cite{Koehler2010} deviations from the Porter-Thomas
distribution for reduced neutron widths of $s$-wave resonances was
revealed. This issue was addressed in \cite{Weiden2010} where the
energy dependance of the widths near a maximum of the neutron
strength function was found to differ from $\sqrt{E}$. In
\cite{Celardo2011} deviations from the PT distribution were seen to
lead naturally from the a careful description of unstable quantum
states with open decay channels. The microscopic physics of
reactions not captured in RMT was shown directly to lead to
deviations from the PT distributions \cite{Volya2011} where the
continuum shell model \cite{PhysRevC.79.044308,PhysRevC.67.054322}
was employed.  It is reasonable to see if the $\Delta_3(L)$
statistic can discriminate between the GOE and a  model which
includes more physics. The shell model with only 2-body interactions
fits the bill. It allows us to get large pure spectra, and include
physical restrictions. The following calculations were carried out
with 12 valence in the $sd$ model space with the  ``USD" interaction
of B. H. Wildenthal using the Oxbash  code. There are 5768 levels
with $J^+T=2^+1$ and 3276 levels with  $J^+T=2^+0$ (see \cite{big}
for more details).  In Fig.\ref{fig:shell} we see the well
established \cite{big} result for $\Delta_3(L)$ for the shell model,
it is well described by RMT. The ${\mathcal N}(\delta)$ is well
within the bounds set by the variance from one spectra to another,
and agrees well with the RMT.

\section{\label{sec:conc}Conclusion}
A maximum likelihood method was devised to gauge the incompleteness
of experimental spectra when a RMT analysis is appropriate. The
method is based on the definition of the $\Delta_3(L)$ statistic.
The distribution of random numbers $\{\delta\}$, the mean of which
is $\Delta_3(L)$, was parameterized. The cumulative distribution
${\mathcal N}(\delta)$  was accurately  fitted with a simple three
parameter function of $\log \delta$: ${\mathcal N}(\delta) =
\frac{1}{2}(1 - \textrm{Erf}[a + b \log \delta + c]$.  These
parameters, $a$, $b,$ and $c$ were parameterized as functions of
$x$, for each $L$, yielding a probability density $p(\delta,x)$ for
$\delta$ with $x$ as a continuous parameter. Our MLM is based on
this $p(\delta,x)$.

The method was tested on a depleted GOE and returned accurate values of $x$. Experimental data was then analyzed. The results for some neutron resonance data sets was consistent with earlier analysis, but occasionally no conclusions could be drawn about the completeness of the data. The acoustic spectra of an aluminum block was then analyzed. The results in 3 out of 6 samples were inconclusive, and for the remaining three, the results were consistent with the conclusions on the experimentalists.

 The expression Eq.\ref{eq:bohigas} was tested and used to gauge $x$. It was found to give a good ensemble average for $x$ but the spread $\sigma_x$ was large. The neutron resonance data and the acoustic data were analyzed with this expression, and agreed most of the time with the MLM results.

The question of intruder levels was addressed and the effects on the $\Delta_3(L)$ statistic, as well as $p(\delta,x)$ were seen to be very similar.

The shell model provides us with  a long pure sequence of pure levels from a system governed by a hamiltonian distinctly different and more physical than those of RMT. Nevertheless, shell model spectra are well described by RMT \cite{big}. We see that the ${\mathcal N}(\delta)$ was no exception, and couldn't distinguish the shell model spectra from the GOE.
\begin{acknowledgments}

We wish to acknowledge the support of the Office of Research Services of the University of Scranton, Vladimir Zelevinsky and Matt Moelter for useful discussions, and the physics department of Temple University who generously accommodated the author for a sabbatical visit. Also we are grateful to the anonymous referee who raised the issue adressed in Sec. \ref{sec:shell}.

\end{acknowledgments}

\bibliography{d3neutronresPRC}

\end{document}